Original Paper

# Tweet for Behavior Change: Using Social Media for the Dissemination of Public Health Messages

Aisling Gough[1], PhD; Ruth F Hunter[2], PhD; Oluwaseun Ajao[3], MSc(Comp Sci); Anna Jurek[3], PhD; Gary McKeown[4], PhD; Jun Hong[3], PhD; Eimear Barrett[2], PhD; Marbeth Ferguson[5], PhD; Gerry McElwee[5], BSc, PGCE, DASE, MSc; Miriam McCarthy[6], MB BcH, MPH; Frank Kee[2], MD

[1]UKCRC Centre of Excellence in Public Health Northern Ireland, School of Medicine, Dentistry & Biomedical Sciences, Queen's University Belfast, Belfast, United Kingdom
[2]UKCRC Centre of Excellence in Public Health Northern Ireland, School of Medicine, Dentistry & Biological Sciences, Queen's University Belfast, Belfast, United Kingdom
[3]School of Electronics, Electrical Engineering & Computer Science, Queen's University Belfast, Belfast, United Kingdom
[4]School of Psychology, Queen's University Belfast, Belfast, United Kingdom
[5]Cancer Focus Northern Ireland, Belfast, United Kingdom
[6]Public Health Agency Northern Ireland, Belfast, United Kingdom

**Corresponding Author:**
Aisling Gough, PhD
UKCRC Centre of Excellence in Public Health Northern Ireland
School of Medicine, Dentistry & Biomedical Sciences
Queen's University Belfast
Institute of Clinical Sciences, Block B
Belfast,
United Kingdom
Phone: 44 028 9097 8931
Fax: 44 028 9023 5900
Email: a.gough@qub.ac.uk

## Abstract

**Background:** Social media public health campaigns have the advantage of tailored messaging at low cost and large reach, but little is known about what would determine their feasibility as tools for inducing attitude and behavior change.

**Objective:** The aim of this study was to test the feasibility of designing, implementing, and evaluating a social media–enabled intervention for skin cancer prevention.

**Methods:** A quasi-experimental feasibility study used social media (Twitter) to disseminate different message "frames" related to care in the sun and cancer prevention. Phase 1 utilized the Northern Ireland cancer charity's Twitter platform (May 1 to July 14, 2015). Following a 2-week "washout" period, Phase 2 commenced (August 1 to September 30, 2015) using a bespoke Twitter platform. Phase 2 also included a Thunderclap, whereby users allowed their social media accounts to automatically post a bespoke message on their behalf. Message frames were categorized into 5 broad categories: humor, shock or disgust, informative, personal stories, and opportunistic. Seed users with a notable following were contacted to be "influencers" in retweeting campaign content. A pre- and postintervention Web-based survey recorded skin cancer prevention knowledge and attitudes in Northern Ireland (population 1.8 million).

**Results:** There were a total of 417,678 tweet impressions, 11,213 engagements, and 1211 retweets related to our campaign. Shocking messages generated the greatest impressions (shock, n=2369; informative, n=2258; humorous, n=1458; story, n=1680), whereas humorous messages generated greater engagement (humorous, n=148; shock, n=147; story, n=117; informative, n=100) and greater engagement rates compared with story tweets. Informative messages, resulted in the greatest number of shares (informative, n=17; humorous, n=10; shock, n=9; story, n=7). The study findings included improved knowledge of skin cancer severity in a pre- and postintervention Web-based survey, with greater awareness that skin cancer is the most common form of cancer (preintervention: 28.4% [95/335] vs postintervention: 39.3% [168/428] answered "True") and that melanoma is most serious (49.1% [165/336] vs 55.5% [238/429]). The results also show improved attitudes toward ultraviolet (UV) exposure and skin cancer with a reduction in agreement that respondents "like to tan" (60.5% [202/334] vs 55.6% [238/428]).

XSL•FO
RenderX

**Conclusions:** Social media–disseminated public health messages reached more than 23% of the Northern Ireland population. A Web-based survey suggested that the campaign might have contributed to improved knowledge and attitudes toward skin cancer among the target population. Findings suggested that shocking and humorous messages generated greatest impressions and engagement, but information-based messages were likely to be shared most. The extent of behavioral change as a result of the campaign remains to be explored, however, the change of attitudes and knowledge is promising. Social media is an inexpensive, effective method for delivering public health messages. However, existing and traditional process evaluation methods may not be suitable for social media.

## Introduction

### Background

Social media is defined as "a group of Internet-based applications that build on the ideological and technological foundations of the Web 2.0, and that allow the creation and exchange of user-generated content" [1]. The considerable rise in the use of social media provides not only an opportunity to reach a large audience [2], but also access to a wealth of user data and the ability to monitor the activities of the audience whom the messages have reached, which will greatly aid our understanding of the underlying mechanisms. Social media statistics from 2015 indicate that 65% of adults are now using social networking sites [3], with more than 310 million monthly active users on Twitter [4] and 1.09 billion daily active users on Facebook [5]. Although largely used by a younger demographic, recent reports point to increased use of Facebook in those 65 years and older [6].

Social media has become ubiquitous, with more people accessing Web-based content by following links on social media than through direct searches [7]. Thus, as a platform used by the public and by health care professionals [8], it presents an ideal opportunity for health promotion. Social media also brings substantial change to the way organizations and individuals can communicate [9-10]. For example, through engaging with social media, the charity Cancer Research UK benefited from a viral social media campaign, the #nomakeupselfie [11]. The charity utilized multiple social media platforms to promote its work, answer questions, and engage in conversations with the public.

We live in a world where, due to the popularity of the smartphone, we have almost instantaneous access to a wealth of specialist information at our fingertips. There is an expectation that health information diffusion will follow suit and health care organizations are turning to social media. For example, Public Health England has responded to the changing landscape of social media and health communication by engaging with digital technologies and switching to an "always on" approach rather than traditional annual campaigns [7].

George et al [12] postulated that social media had direct public health relevance because social networks could have an important influence on health behaviors and outcomes. However, public health agencies have not yet harnessed the full potential of social media [13-14]. Chou et al [14] particularly noted the need for public health interventions to "harness the participatory nature of social media." Heldman et al [15] proposed that public health organizations and practitioners too often used social media for the traditional 1-way broadcast of information, rather than utilizing the opportunity to engage audiences in 2-way communications, or as they call it, being "truly social."

There is a wealth of opportunity to use social media for health promotion, through targeted messages, the ability to interact with the public, target hard-to-reach groups, and create dynamic campaigns [12,13,16-18]. Pagoto et al [19] alluded to the ability to be "in the participant's pocket" through social media providing advice and support. Opportunities for discussion (social connection) are considered to be 14 times more effective with social media compared with the written word [20], with reports that information shared via social media resulted in greater knowledge scores than when shared via pamphlets [21]. With this lies the potential of social media to overcome barriers with regard to access to information [22] and literacy. Social media has, in essence, flattened the world with regard to health information, providing potential for building bridges between disconnected groups.

Despite a recent review alluding to a positive effect of social network interventions on health behavior-related outcomes [23], studies of social media as a channel for health promotion are limited [18]. Although social media is being increasingly used by public health departments, from a research perspective, it is not yet clear how best to capitalize on social media for raising awareness and, ultimately, triggering behavioral change. Research is lacking with regard to developing and implementing such campaigns. Nor do we know what a successful campaign entails, be that (as some have suggested) the number of followers of the campaign social media platforms, the number of retweets or shares of a given message, or simply the number of people who see a given message. It has been proposed that through surveillance of Twitter, such data can be used as a proxy measure of the success or effectiveness of a given health message or public health campaign [24]. However, we still find ourselves asking, "What makes a social media campaign successful?" Do shares or "likes" imply behavior change? In the marketing sector, it may be clearer with regard to increased sales or website clicks, but in the realm of public health, such questions remain unanswered.

As such, there have been calls for more research to focus on social media and communication technologies [25]. Given the number of unanswered questions around the feasibility of using social media for health promotion and public health, this study aims to address some of these through reporting the findings of a mass communication Twitter campaign for the prevention of skin cancer.

### Aim

The aim of this study was to test the feasibility of designing, implementing, and evaluating a bespoke social media-enabled intervention for the dissemination of public health messages to prevent skin cancer.

### Research Objectives

This mixed-methods study investigated the feasibility of implementing a social media–enabled public health campaign focusing on skin cancer to increase knowledge and attitudes toward care in the sun. The research had the following objectives:

- To investigate the feasibility of a bespoke social media-enabled campaign on skin cancer attitudes and knowledge
- To investigate the impact of employing different message frames on social media
- To investigate whether there are benefits to using promoted messages, influencers, and a Thunderclap for the diffusion of messages on social media
- To determine the appropriate process evaluation measures and access to data for a social media campaign (user demographic details including gender)
- To investigate whether there is an appropriate control group for a social media campaign

## Methods

### Why a Skin Cancer Campaign?

Skin cancer is the most common form of cancer diagnosed in Northern Ireland, with more than 4000 cases diagnosed annually [26-27]. In Australia, campaigns such as "Slip, Slap, Slop" have been run for more than two decades. Such campaigns have increased skin cancer awareness and sun-safe behaviors [28]. Nationally, Cancer Research UK have developed the "SunSmart" campaign [29], which focused on raising awareness on skin cancer through skin protection, avoiding sunburn and use of sunscreens. Regionally, the leading cancer charity in partnership with the Public Health Agency has coordinated the "Care in the Sun" campaign (which is similar in many respects to SunSmart). This study was conducted to assess baseline and post campaign levels of sun-safe knowledge, attitudes, and behavior.

To establish the baseline parameters for the campaign, we utilized a household survey based on the questions used in the SunSmart omnibus survey. A postcode stratified sample of 750 was selected based on a representative distribution across Northern Ireland. The results from the household survey demonstrated that although the majority of respondents were aware that sun exposure could cause skin cancer (80.7%, 605/750), and aware that skin cancer could lead to death (88.9%, 667/750), few were aware that skin cancer was the most common cancer and that melanoma was the most serious type (41.1%, 308/750, answered "Don't Know"). Almost 50% of participants considered a suntan to look healthy (49.2%, 369/750) and fewer than 10% reported frequent skin checks (6.4%, 48/750). This knowledge of the known gaps in sun-safe attitudes pertaining to skin cancer evidence was the motivation for the regional Public Health Agency identifying skin cancer as a priority area for its social media campaign.

### Why Twitter?

Twitter was selected as the social media platform for diffusing our campaign messages, as Twitter information is posted voluntarily and is in the public domain. Unlike other social media platforms, Twitter provides several application programming interfaces (APIs) that allow real-time access to vast amounts of content, thereby aiding our understanding of social media processes. *Adoreboard*, a University spin-out company, enabled access to Twitter streaming data, which are preprocessed to minimize "noise," and allow maximal recovery of textual information and user metadata. Thus, the captured data are cleaned by removing unwanted messages and irrelevant tweets, which constitute noise in the message corpus. We aimed to remove tweets that did not contain the relevant hashtags of the campaign. The data cleaning process was initially simplified by the use of unique hashtags, and included the removal of blank tweets and spam tweets posted for promotion of a product or service or those automatically broadcast by robots. Preprocessing of the data still remains paramount. Thus, the setting of the message filters on the Twitter stream ensures that only the required messages are captured and analyzed.

### Design

A quasi-experimental feasibility study—specifically an interrupted time series with comparison design—was implemented to assess the efficacy of the social media intervention. A "cross-over" design was utilized, whereby the regional cancer charity's Twitter account hosted the campaign between May 1 and July 14, 2015 (Phase 1), followed by a gap of 2 weeks ("washout"), and then a phase of campaign messages posted from a new social media account between August 1 and September 30, 2015 (Phase 2). The 2 intervention phases were differentiated on the basis of the host platform to establish whether starting a new social media account would impact on message diffusion in comparison to using an already established social media account of a local cancer advocacy charity. Phase 1 was longer in duration (by 2 weeks) to account for any reduced social media interactions due to a national holiday period in early July. The protocol was developed in accordance with the CONSORT-EHEALTH checklist [30].

### Control Group

Social media analytics were tracked in 2 geographical areas, through geo-location information contained within a subset of tweets: 1 area exposed to the campaign (Northern Ireland), and another area that did not receive the specific elements of the campaign; this was used as a control area for comparison (Wales). The volume of tweets related to a list of predefined keyword search terms (Multimedia Appendix 1) was compared pre-and postcampaign in order to track the messages in each location.

## Intervention Development and Implementation

A detailed description of the intervention design can be found in Multimedia Appendix 2. During Phase 1 of the campaign (May 1 to July 14, 2015), we utilized the existing regional cancer charity Twitter account. Each week, the different message frames (informative, story, shock, humor, and a final opportunistic or responsive category) were utilized (Figure 1). Messages were focused on both skin surveillance and general care in the sun and skin cancer prevention. Seed user and opportunistic messages were utilized where appropriate.

Phase 2 of the campaign utilized a bespoke Twitter account to disseminate messages. Similar content was used for the second phase as in the first, which included both skin surveillance and general care in the sun or cancer prevention. Phase 2 also included a Thunderclap, a Web-based "flash-mob" of messages involving users to permit their social media accounts to automatically post a common message, related to the campaign, on their behalf. The Thunderclap took place on midday of September 1, 2015.

In both phases of the campaign, paid-for promoted posts on Twitter, to the value of £10, were used to enhance Web-based content by increasing the number of people who saw the messages. Promoted tweets work on a "cost per click" basis, whereby an allocated budget is set by the user (eg, £10) and that tweet is promoted to the specified audience until the budget runs out. Audiences for the promoted posts were specifically targeted to those living in Northern Ireland and aged 18 years or older.

**Figure 1.** Intervention timeline.

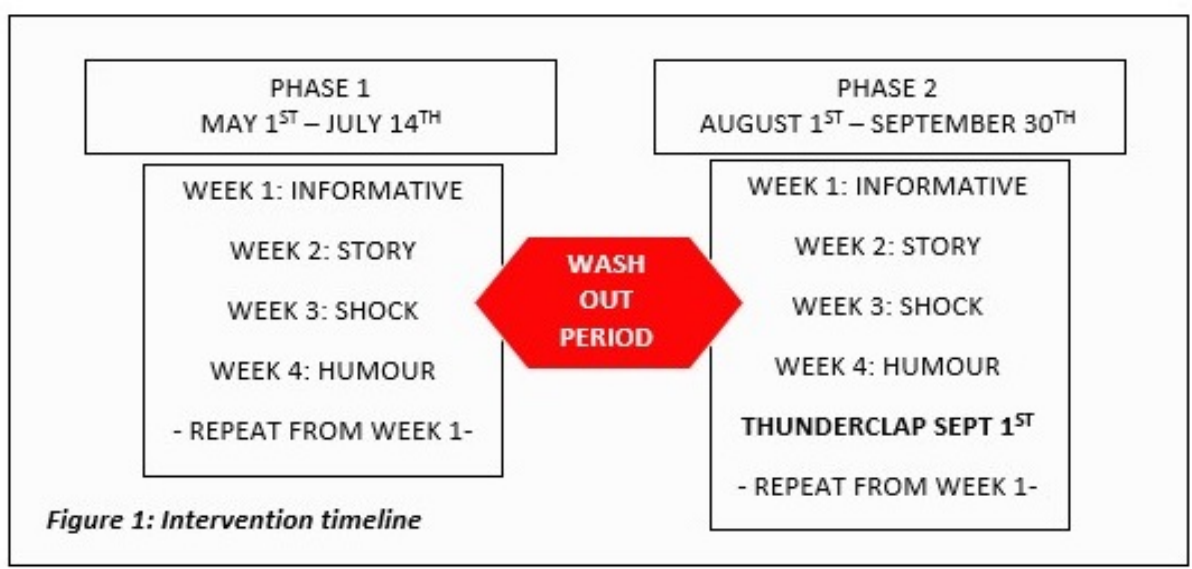

Figure 1: Intervention timeline

## Measures

Twitter analytics for key search terms (Multimedia Appendix 1) related to skin cancer and care in the sun were collected both before and after the campaign (April 2015 and October 2015) to serve as a comparison with the control group to establish whether the campaign resulted in greater use of such search terms, in Northern Ireland, following the campaign. Throughout the campaign, social media analytics were collected from Twitter dashboards. Access to Twitter streaming data was also enabled by *Adoreboard*, a University spin-out company. The most commonly cited and readily available social media metrics collected are defined and described later [17,24,31]. Such metrics may tell us the extent to which a message spreads by detailing the number of users who see it, who respond to it, or who subsequently share that message with their friends or followers.

Impressions: The number of views of a particular post from users who saw it appearing on their timeline or through search results.

Engagements: The number of clicks on the message, the picture posted, or the number of people who actively engaged with a post including likes, comments, shares, and retweets.

Engagement rate: The ratio of engagements to impressions.

*Likes*: Posts can be endorsed by the friends or followers of users that post messages by "liking" them (alternatively known as "favorites" on Twitter).

Shares: Similarly readers of a post or status update who found the message interesting could also rebroadcast them by simply sharing (or retweeting) them.

Data related to social media user demographics are limited. Twitter provides limited public information about the profile of its users within the description field. Typically this does not include gender, but we subsequently aimed to infer the gender of the participants in the study based on their given names on Twitter.

## Pre- and Postintervention Web-Based Survey

The Checklist for Reporting Results of Internet E-Surveys (CHERRIES) checklist for the reporting of Web-based surveys [32] was taken into account in this study (Multimedia Appendix 3). An advertisement (Multimedia Appendix 4) was placed on social media in April 2015, and again in October 2015, inviting adults aged >18 years in Northern Ireland to participate in a survey for the chance to win an iPad Mini. Those who clicked on the advertisement were redirected to a Qualtrics survey website modeled on the Cancer Research UK SunSmart survey [29]. The survey took approximately 15-20 minutes to complete. Paid-for "promoted" tweets were used to reach a wider audience. In line with recommendations from the regional cancer charity, adverts were promoted to the value of £15 on Facebook and £10 on Twitter. The sample was stratified by age and region. As an example of reach of promoted posts on Facebook, a £15 limit has the potential to reach 770-2000 people living in Northern Ireland aged 18-65+ years. The surveys consisted of 37 multiple-choice questions subdivided into 3 broad subsections: sociodemographic information; skin cancer prevention; and psycho-social mediators of behavioral change. Differences between the preintervention period and postintervention period served as an assessment of the impact of the intervention. The primary outcomes were change in sun protection attitudes and knowledge regarding skin cancer. Completed surveys, as indicated by completion of the final question, were included for analyses. IP addresses were not checked for duplicate users.

### Data Analysis

Focus group and workshop discussions were audio-recorded and transcribed verbatim and anonymized. Transcripts were read repeatedly, initial codes identified, and themes collated and analyzed using an "a priori" thematic "Framework" method to produce themes related to perspectives of professionals and users [33].

Data were compared for the pre- and postintervention survey, including social media usage, demographics, and knowledge and attitudes toward UV exposure and skin cancer prevention. Descriptive statistics (frequencies) of responses to questions were tabulated, and cross-tabulations used to report responses to questions by gender, age, and other sociodemographic characteristics. Tests of significance were omitted due to the nature of the study and the appropriateness of applying such tests to feasibility studies [34].

Social media analytics were gathered for each post, for each message frame and the overall campaign, and frequencies utilized to compare impressions, engagement, and shares for message frames.

### Measuring Twitter Analytics

Twitter data, collected via a service provider (Adoreboard), allowed greater access to the Twitter Firehose compared with the public APIs. The retrieved tweets and metadata were used to tabulate information such as the impressions and engagements of each tweet and to establish the frequencies of hashtag use and message spread (retweets). Metadata fields were also searched for relevant search terms (Multimedia Appendix 1) for the pre- and postcampaign frequency comparison.

## Results

### Can a Bespoke Social Media Campaign on Skin Cancer Impact on Attitudes and Knowledge?

*Demographic Characteristics*

A total of 337 participants completed the precampaign Web-based survey, compared with 429 who completed the postintervention Web-based survey (Table 1). The age distribution of participants both pre- and postcampaign was similar, with 41% of each aged 18-29 years, and respondents were more likely to be females (84.6% [281/337] preintervention; 80.4% [345/429] postintervention). Marital status and educational attainment distributions were also similar pre- and postcampaign, with more than half respondents reporting a University degree or higher (preintervention 54.6% [184/339] vs postintervention 51.5% [221/429]). More of the preintervention households reported an annual income greater than £20,001 (preintervention 52.8% [158/300] vs postintervention 40.4% [152/376]). A substantial proportion of respondents lived in Belfast (preintervention 41.9% vs postintervention 35.4%). About 15.5% of the general population of Northern Ireland lives in Belfast according to the Northern Ireland census [35].

As reported by the 2011 Northern Ireland Census [35], the population of Northern Ireland has 51% females and 49% males; thus, a greater number of females responded to both the pre- and postcampaign [36]. Campaign respondents were more educated than the Northern Ireland population (29% aged 16+ years had no qualifications) and were of a similar age (2011 Northern Ireland Census median age, 37 years).

**Table 1.** Respondent demographic characteristics in the pre- and postintervention Web-based surveys.

| Demographic characteristics | | Precampaign (n=337) | Postcampaign (n=429) |
|---|---|---|---|
| Age (year) | | 35.6 | 35.2 |
| **Gender, n (%)** | | | |
| | Male | 51 (15.4) | 84 (19.6) |
| | Female | 281 (84.6) | 345 (80.4) |
| **Marital status, n (%)** | | n=33 | n=429 |
| | Single | 150 (44.5) | 182 (42.4) |
| | Married or cohabiting | 161 (47.8) | 224 (52.2) |
| | Divorced or widowed | 26 (7.7) | 23 (5.4) |
| **Highest level of education, n (%)** | | n=339 | n=429 |
| | None | 13 (3.9) | 11 (2.6) |
| | GCSE or equivalent | 37 (11.0) | 41 (9.6) |
| | "A" level or equivalent | 103 (30.6) | 156 (36.4) |
| | Degree or higher | 184 (54.6) | 221 (51.5) |
| **Housing tenure, n (%)** | | n=336 | n=429 |
| | Rent or other | 134 (39.9) | 162 (37.8) |
| | Mortgage or co-ownership | 150 (44.6) | 189 (44.1) |
| | Owned outright | 52 (15.5) | 78 (18.2) |
| **Annual income, n (%)** | | n=300 | n=376 |
| | <£12,000 | 77 (25.8) | 105 (27.9) |
| | £12,001 to £20,000 | 64 (21.4) | 119 (31.6) |
| | >£20,001 | 158 (52.8) | 152 (40.4) |

### Attitudes to UV Exposure and Skin Cancer Prevention

Postcampaign, there was a trend toward improved attitudes toward UV exposure and skin cancer with a reduction in agreement that respondents "like to tan" (pre- 60.5% [202/334] vs postcampaign 55.6% [238/428]), that "a tanned person looks more healthy" (55.9% [186/333] vs 52.7% [225/427]) or attractive (48.6% [162/333] vs 43.7% [186/426]). The postcampaign also noted a trend toward improved attitude to UV exposure, with greater agreement that protection from the sun can help avoid skin cancer (62.6% [209/334] vs 65.0% [278/428]). Trends in change of care in the sun attitudes are shown in Table 2.

**Table 2.** Trends in change of care in the sun attitude and knowledge.

| Trends | | Precampaign | Postcampaign |
|---|---|---|---|
| Care in the sun attitude | | | |
| **I like to tan, n (%)** | | n=334 | n=428 |
| | Agree | 186 (60.5) | 238 (55.6) |
| | Neutral | 66 (19.8) | 98 (22.9) |
| | Disagree | 66 (19.8) | 92 (21.5) |
| **A suntanned person looks more healthy, n (%)** | | n=333 | n=427 |
| | Agree | 186 (55.9) | 225 (52.7) |
| | Neutral | 74 (22.2) | 102 (23.9) |
| | Disagree | 73 (21.9) | 100 (23.4) |
| **If I protect myself from the sun I can avoid skin cancer, n (%)** | | n=334 | n=428 |
| | Agree | 209 (62.6) | 278 (65.0) |
| | Neutral | 41 (12.3) | 68 (15.9) |
| | Disagree | 84 (25.1) | 82 (19.2) |
| Care in the sun knowledge | | | |
| **Sun exposure causes most skin cancers, n (%)** | | n=336 | n=428 |
| | True | 269 (80.1) | 346 (80.8) |
| | Don't Know | 46 (13.7) | 49 (11.4) |
| | False | 21 (6.3) | 33 (7.7) |
| **Skin cancer is the most common form of cancer, n (%)** | | n=335 | n=428 |
| | True | 95 (28.4) | 168 (39.3) |
| | Don't Know | 133 (39.7) | 166 (38.8) |
| | False | 107 (31.9) | 94 (22.0) |
| **Melanoma is the least serious form of skin cancer, n (%)** | | n=336 | n=429 |
| | True | 35 (10.4) | 36 (8.4) |
| | Don't Know | 136 (40.5) | 155 (36.1) |
| | False | 165 (49.4) | 238 (55.5) |

### Knowledge of Skin Cancer Prevention

Postintervention there was a trend toward improved knowledge of skin cancer prevention (Table 2), with greater awareness that skin cancer is the most common form of cancer (28.4% [95/335] vs 39.3% [168/428] answered "True") and that melanoma is most serious (49.1% [165/336] vs 55.5% [238/429]). There was also a trend toward improved awareness that sun's rays are strongest at midday (91.3% [306/335] vs 93.5% [400/428]) and that people with fair-colored skin require the most protection (73.8% [248/336] vs 77.6% [332/428]).

### Campaign Reach

#### Influence of Message Frames on Social Media

Of the 4 specific message frames utilized (informative; personal story; shock or disgust; humorous), a shock or disgust tweet (#eek) achieved the most impressions (n=2369), followed by an informative message (#info, n=2258; Table 3). The most engaging tweet was that with humor (#geg, n=148), followed by 1 characterized by shock or disgust (#eek; n=147). The most retweeted message was that of an informative nature (#info), shared by 17 followers. The most retweeted #story message was shared 7 times, compared with 9 for #eek and 10 for #geg messages. When comparing the median values for each message frame, shocking messages achieved greater impressions (median 565), engagements (15.5), and retweets (2.5), whereas humor messages achieved a greater median engagement rate (2.5%).

### Are There Benefits to Using Promoted messages, Influencers, and a Thunderclap for the Diffusion of Messages on Social Media?

#### Influencers

Tweets that included an influencer in the message generated greater numbers of impressions. Influencer posts also created the most impressions when on a #eek post (n=11,349) and a #story post (n=9612). Tweets that were paid-for—promoted posts—did not notably increase impressions, engagements, or retweets. Table 3 shows the top Twitter analytics for each message frame.

*Opportunistic Messages*

Considering messages that were of an opportunistic nature, the greatest number of impressions for a single message was 2993, whereas the greatest number of engagements on an opportunistic tweet was 103. The greatest number of retweets for an opportunistic message was 8.

*Thunderclap*

The campaign Thunderclap had a minimum goal of 100 supporters (in order for the Thunderclap to activate), and achieved a total of 122 supporters and social reach of 454,207 (sum total of the friends and followers of campaign supporters). Top tweets mentioning or encouraging support for the Thunderclap achieved 2527 impressions, 2 engagements, and 7 retweets. Thunderclap-related tweets, which included an influencer in the tweet, achieved greater numbers of impressions (n=11,740) than Thunderclap tweets that did not (n=2527).

## What Are Appropriate Process Evaluation Measures?

*Twitter Analytics*

During the campaign period, there was a total of 417,678 tweet impressions based on the campaign (Table 3). Post engagements reached 11,213, and there was a total of 1211 retweets. Of these, 92 retweets were part of the Thunderclap. A single tweet achieved 11,349 impressions. The same tweet was the most engaging, resulting in 811 engagements. The most retweets on any 1 post was 17.

**Table 3.** Twitter analytic attributes for message frames.

| Message frames | | Impressions (median) | Engagement (median) | Engagement rate in % (median) | Retweet (median) |
| --- | --- | --- | --- | --- | --- |
| **All tweets** | | | | | |
| | All tweets | 6367 | 196 | 14.8 | 17 |
| | + Influencer | 11349 | 811 | 12.0 | 13 |
| | + Promoted | 4808 | 304 | 11.5 | 12 |
| **Info (informative)** | | | | | |
| | Info | 2258 (443.0) | 100 (8.0) | 8.3 (2) | 17 (2) |
| | + Influencer | 3161 | 106 | 8.1 | 7 |
| | + Promoted | 2335 | 98 | 4.2 | 13 |
| **Story (personal story)** | | | | | |
| | Story | 1680 (390.5) | 117 (6.5) | 3.6 (1.3) | 7 (1) |
| | + Influencer | 9612 | 52 | 3.0 | 5 |
| | + Promoted | 1210 | 19 | 7.0 | 3 |
| **Eek (shock or disgust)** | | | | | |
| | Eek | 2369 (565.0) | 147 (15.5) | 10.1 (2.2) | 9 (2.5) |
| | + Influencer | 11349 | 811 | 7.1 | 11 |
| | + Promoted | 2655 | 301 | 11.5 | 5 |
| **Geg (humorous)** | | | | | |
| | Geg | 1458 (487.0) | 148 (12.0) | 14.8 (2.5) | 10 (2) |
| | + Influencer | 1459 | 21 | 5.7 | 2 |
| | + Promoted | 4808 | 67 | 1.4 | 11 |
| **Opportunistic** | | | | | |
| | Opportunistic | 2993 (385) | 103 (5) | 8.40 (1.5) | 8 (1) |
| | + Influencer | 10674 | 211 | 12.00 | 12 |
| | + Promoted | 6367 | 196 | 4.60 | 14 |
| **Influencer** | | | | | |
| | Influencer | 11349 | 811 | 12.00 | 11 |
| | +Promoted | 2110 | 76 | 11.50 | 4 |
| **Thunderclap** | | | | | |
| | Thunder | 2527 | 2 | 4.80 | 7 |
| | + Influencer | 11740 | 3 | 7.80 | 6 |
| | +Promoted | 135 | 3 | 2.20 | 0 |

### Is There an Appropriate Control Group for a Social Media Campaign?

A search for keywords relating to sun exposure and skin cancer, geo-tagged to NI, returned 15,964 and 14,168 tweets for April and October 2015, respectively (Multimedia Appendix 1). For Wales, 50,164 and 51,634 tweets were returned for April and October 2015, respectively (The population of Northern Ireland is 1.8 million, while that of Wales is 3.0 million). Comparing the total key words retrieved for an Northern Ireland geo-located word search with that of Wales in the pre- and postcampaign period, postcampaign there was an increase in those geo-located to Wales whereas there was a decrease in those geo-located to NI. However, our designated campaign hashtags did not appear among the key word search retrieved from Wales.

## Discussion

### Principal Findings

The aim of this study was to develop, implement, and evaluate a social media public health campaign. In doing so, we sought to uncover the feasibility of using social media (Twitter) for the dissemination of public health messages, to investigate the impact and appropriateness of different message frames, promotion techniques, and evaluation measures. Our findings suggested that social media was indeed a feasible platform for the delivery of a public health campaign.

### Investigating the Impact of a Bespoke Social Media Campaign on Skin Cancer Attitudes and Knowledge

Social media is a feasible platform for the dissemination of public health messages owing to the ability to provide dynamic and tailored messages to an audience in real time. The results from the pre- and postcampaign Web-based survey showed a trend toward improvements in both knowledge and attitudes with improved awareness that sun protection can reduce skin cancer risk and greater awareness of the severity of skin cancer. The results of Web-based surveys have to be interpreted cautiously [37-38], as they cannot serve as accurate indices of overall population knowledge about public health issues. Nevertheless there is probably greater correspondence between the characteristics of respondents to Web-based surveys and those of social media users, who are the natural target of a social media campaign, than is the case with traditional respondents to face-to-face household surveys.

### Investigating the Impact of Employing Different Message Frames on Social Media

Message frames with shocking (#eek) content generated the greatest number of impressions, whereas humorous (#geg) messages resulted in greater public engagement on social media, compared with personal story messages. Message framing on social media has been the topic of much debate in the research literature. The idea of gain-framed versus loss-framed messages to encourage illness prevention behaviors has been explored in relation to skin cancer prevention, with mixed results. Gallagher and Updegraff [39] found that gain-framed messages were more likely than loss-framed messages to encourage skin cancer prevention behaviors, whereas others found no difference [40]. Moreover, graphic warning messages, like some used in this study, have been shown to be more effective in promoting behavioral change, particularly related to tobacco use, as they attract attention and evoke emotion and motivation to quit [41,42]. For example, some work has employed graphic content and message framing for skin cancer prevention and indoor tanning [42] while others [43] appropriated negative emotions to generate discussion. It has been postulated that exposure to negative emotions may affect risk perceptions and thus talking about them can serve as a means of dealing with such emotions [43,44].

Previous work has highlighted the role of fear-based approaches related to public health campaigns [45-46] in raising awareness by attracting attention, much like this study, which found that more impressions were evident from "shocking" messages tweeted. With fear-based approaches too comes the possibility of unintended effects such as dissonance or desensitization, as alluded to by Cho and Salmon [47]. However, a recent work from Bail [48] found that positive emotional content increased the virility of messages used for a social media Facebook campaign. Bail [48] suggests that social media campaigns must not rely on fear-based tactics to draw attention to their cause, but that campaigns may benefit from the use of positive emotional language. The use of humor in public health messages for behavioral change has been explored [31. Through use of a "Laugh Model," the authors sought to learn from business, marketing, and branding strategies in order to prioritize the use of humor and entertainment in health promotion messages. They implemented a social media campaign to promote healthy family meals in the Utah region, with humor and entertainment underpinning the campaign. The campaign was deemed to be successful in reaching 10%-12% of the target population, achieving 17,377 Facebook impressions, 28,800 Twitter impressions, and 5591 Web-based engagements. The population of Utah is 2.9 million, and their Twitter impression and engagement rates were thus 0.01/100,000 and 0.002/100,000, respectively, compared with 0.22/100,000 and 0.006/100,000 in this campaign. The authors found their humorous posts to be most successful, with an engagement rate of 9.7%, suggesting that such engaging techniques might be more effective than educational techniques. In line with the Laugh Model [31], a humorous message in this study achieved the greatest engagement rate, reaching 14.8%.

Twitter users have a variety of motivations for re-Tweeting. Ramdhani [49] suggested that the motivating factors included self-enhancement, social interaction, personal benefit and appreciation, and (through humor), entertainment. Ramdhani [49] also noted that providing information was of little importance as a motivating factor. However, in this study, the most retweeted message was that of an informative nature. Kandadai et al [50] noted that users were selective in determining what to retweet, and when the information was deemed valuable and credible, it was shared. The psychology of sharing has recently been explored in a study among 2500 Web-based users of the New York Times [51], which found that the most common reasons for people to share material with others across social media were to provide entertainment (94%), spread the word about a cause or issue they care about (84%), as a method of "information management" allowing them to process information more thoroughly when sharing it (73%), and self-fulfillment (69%). Future qualitative work would be required to tease out the motives of social media followers in choosing which health messages to retweet. Moreover, it would be advantageous for message types to be rated by social media users in order to ensure consistency with regard to categorization assigned by the research team.

## Investigating Whether There Are Benefits to Using Promoted Messages, Influencers, and a Thunderclap for the Diffusion of Messages on Social Media

### Promoted Messages

Although Lister et al [31] used paid posts to increase the number of followers and website traffic, this study did not find value in the use of paid-for, promoted tweets. Actually overall, promoted posts resulted in fewer impressions and retweets than both "organic posts" and those that included an influencer. However, promoted posts had the potential advantage of targeting specific groups, by location, age, gender, interest, and so forth. As this study had quite a broad target audience of adults (+18 years) living in Northern Ireland, future studies and interventions with a more specific target may see greater benefit from using promoted posts.

### Influencers

Based on the feedback from the focus groups and "co-design" workshops, we included the use of "influencers" or seeds and a unique hashtag for the social media campaign. By doing so, there was suggestive evidence in this study that the number of impressions and engagements was greater where influencers were utilized (Table 3). This was in line with the recommendations from the Social Bakers blog ("What we can learn from the top 2 Twitter accounts") [52], which included working with viral influencers, the use of a unique hashtag. It is not surprising that greater impressions result from influencer posts because they tend to have large numbers of followers. However, following Ramdhani's [49] and Bret's [51] findings related to motives for tweeting, the selection of influencers is of utmost importance. Thus, rather than selecting influencers based on their large following, to drive post impressions, further thought should be given to the influencer's social media "tone of voice" and whether the intervention messages are in line with the influencer's and their following. Ultimately this may help a given message reach an audience primed to engage with the content.

The problem of identifying the most influential users in social networks has been considered by many studies. The most common factors that have been considered as indicators of user's influence in social media are the number of followers, the number of friends, the number of days that the user exists on Twitter, the number of tweets posted by the user in the past, or the number of times the user was mentioned in the past [53]. More empirical research is required to measure the influence of a user based on his past activities [54] (ie, how many users he influenced in the past conversations).

The purpose of using influential users in the campaign was to increase the impact of the messages, so that more people might engage with the conversation. In social media the impact of a message is determined by how well the message propagates in the network. This is commonly referred to as information diffusion. A number of ways of assessing information diffusion has been considered in different studies. The most common way of quantifying the diffusion of a message in a network is through the volume of users influenced by the message [55-58]. Influenced users (often referred to as activated users) are those who engaged with a post through liking, commenting, sharing, or retweeting. Studies have evaluated the diffusion of a tweet through its retweetability (ie, the length of the retweet chain) [55,58]. According to Wang et al [59], a message tends to propagate better if not only your friends, but also friends of friends, are getting involved in the conversation. Therefore, the number of friendship hoops that a message has traveled was taken under consideration while assessing the information diffusion.

Lots of work has been focused on the challenge of predicting diffusion of a message in social media. The relation between the influence of a user and the information diffusion has been investigated in a number of studies. The factor that is most commonly applied to estimate the level of user's influence is the number of his or her followers. It has been demonstrated that there is a correlation between the number of followers and the length of the retweet chain [55]. The study by Yang and Counts [56] suggested that the number of times a user has been mentioned in the past is a good predicator of the number of his or her followers who might be influenced. Other studies have focused on developing predictive models for information diffusion using machine learning-based approaches. Naveed et al [58], for example, built a predictive retweet model using logistic regression. They used some of the aforementioned user-related features with additional features related to the content of the tweets (eg, whether the message contains a uniform resource locator, hashtag, or mention). Hong et al addressed the problem of predicting the popularity of a tweet (ie, number of retweets) by formulating it as a classification problem [60]. Instead of predicting the exact number of retweets, each tweet was assigned to a category representing an estimated volume of retweets. Another approach to modeling the information diffusion was presented by Yang and Leskovec [61], where the number of newly influenced users was modeled as a function of which other users were influenced in the past. Wang et al [59] proposed an alternative model that was able to predict the density of influenced users over time based on how well the message spread in the early phase.

These machine learning approaches highlight a number of interesting ideas that can be applied in future social media campaigns, suggesting that there are methods that can be used to automate and enhance campaign assessment processes. Both identifying the most influential users in the network and predicting the propagation of messages could be used to increase the impact of a social media campaign. The results of this study indicate that using influencers as seeds increases the number of impressions and engagements. At the same time, we noticed that the level of influence differed among different seed users. Therefore, it could be beneficial to consider factors other than the number of followers while selecting the seed users for the campaign. Using some of the predictive models described earlier could help in the assessment of the propagation of the messages that could be used as the predicator of the campaign's impact.

### Thunderclap

The Thunderclap campaign exceeded its target of reaching 100 supporters, (and achieved a total of 122 supporters). This target was somewhat arbitrary but exceeded the number achieved by previous campaigns of the Regional Public Health Agency. The Thunderclap required users to pledge their support and thus allow a bespoke campaign message to be posted from their chosen social media account, resulting in widespread social reach with more than 450,000 people seeing the campaign message. Thus, a Thunderclap is a useful tool for spreading awareness provided it is utilized correctly, and adequately promoted and explained prior to launch. Thus, it is important to ensure awareness among users and actively pledge support by following the Thunderclap link, rather than simply retweeting or "liking" the message advertising the Thunderclap.

The scheduling of campaign messages was informed by both the focus groups and availability of the host's (a regional cancer charity) social media accounts. Messages were posted between 3 and 4 days per week on Twitter, with the same message (or minor variations of the same message) posted up to 4 times in a day at different times. Moz Blog [62], in 2012, reported that the average lifespan of a tweet was 18 minutes, for accounts with fewer than 1000 followers, and that although retweets extended the lifespan, most retweets happened in the first 7 minutes of a message being posted. Such detailed analysis was beyond the scope of this study; however, we did observe that the greatest lifespan of 1 of our tweets was 64 days (ie, there was a retweet 64 days following the original tweet). Increasing numbers of social media marketing tools have appeared in recent years, and future work may benefit from utilizing such applications. Such tools include Twitalyzer for Twitter or Likealyzer for Facebook, which offer more than is available from the traditional social media platform dashboard analytics, including recommendations for the best times to publish social media posts, whether users respond more to photos or videos, and ranking comparisons to similar social media profiles. Future work should therefore attempt to capitalize on such resources. Moreover, in this study, it may be possible that greater tweeting frequency of content further instilled the message to users or provided opportunity to reach different social media users at various time points. However, on the contrary, the increased volume of content may, as has been suggested [63], inadvertently decrease the perceived importance of the content, particularly as "shares" were few in this study. Thus, a "less is more" approach may be beneficial if the aim is to achieve shareable content and subsequent message diffusion.

### Determining the Appropriate Process Evaluation Measures and Access to Data for a Social Media Campaign

This study utilized commonly cited and most readily available Twitter analytics to evaluate the campaign: impressions, engagements, likes, and shares. Although such markers are commonly used in the literature [17,24,31], one might contest their appropriateness as evaluation measures for a public health campaign [64]. Although measures such as impressions are useful in determining how many users see a given message, and retweets in determining the number who share such messages with friends or followers, it may not be appropriate to infer specific meanings from such actions. Does liking or retweeting a message infer that the user supports the campaign message or wants their friends to be aware of such advice or will indeed take heed of the message and act on such advice—or example in our case—to apply sunscreen? It would be naive to infer that they are good barometers of impending behavioral change.

Thus, with calls specifically for eHealth interventions across the board and particularly with regard to melanoma [65], new research methods for social media are needed, perhaps through the adaptation of traditional methods. For example, this study delivered its Web-based survey via social media. Moreover, applications of traditional methods (to inform the design of search terms) for social media are beginning to emerge, such as the use of netnography, a fusion of ethnography with Internet analytics [66]. Crowdsourcing and photo and video elicitation techniques may also be adapted for social media to gain a deeper understanding of perceptions, attitudes, and behavioral change.

### Investigating Whether There Is an Appropriate "Control Group" for a Social Media Campaign

This feasibility study was unable to ascertain an appropriate "control group" for a social media–enabled public health intervention. Wales was chosen as a "control group" for the campaign. However, with the use of our specific hashtags, we did not expect any impact or social media footprint in Wales and we found virtually none. For a campaign based around behavior in the sun, clearly local geography and weather conditions are likely to have an impact on message reach and engagement. Although there are bound to be very local variations, Northern Ireland and Wales "enjoy" broadly the same weather, and so we anticipated that social media traffic in Wales could tell us something about the background influence of these weather effects. Moreover, although weather data was collected for Northern Ireland throughout the campaign period, stronger associations between weather and retweets emerged during Phase 2 of the campaign. This was at a time frame that was closer to the "peak" of summer in Northern Ireland. However, during the summer of 2015, Northern Ireland experienced one of the coldest, wettest summers in approximately 30 years, and so any conclusions must be tentative.

Although this study identified the appropriateness of social media for a public health campaign, the challenge is to find how to transfer traditional evaluation principles into the world of social media. Innovative methods are emerging with regard to social media. An instrumental variable approach to study happiness and weather effects has also been reported [67]. Techniques are also emerging to better measure and assess the effects on sentiment in social media through the use of "emoji" [68]. With few studies having examined the success of social media to promote knowledge and adoption of health behaviors [69], there is room for methodological innovation because traditional randomized controlled trial methods and process evaluation measures (MRC Guidance) have little to say on social media interventions.

This study sought guidance from a statistician to determine the best course of action for handling Twitter data. Traditional statistical analyses may not be appropriate given the clustered and dependent nature of tweets. The results of this study should be considered within the context of other limitations. Although a shocking tweet (#eek) achieved the greatest number of impressions, this was likely driven by the associated "influencer" because median impressions for tweets were substantially less. Nonetheless, taking median values, shocking tweets achieved most impressions. Moreover, tweet content was determined by the research team and verified by participants at a codesign workshop. Future work would benefit from an assessment of content agreement when determining message frames applied to tweets. For example, what was considered to be a humorous message by the research team may be deemed as shocking to another user.

The National Institute for Health and Care Excellence guidelines for sunlight exposure [70] were updated after the completion of this intervention, and its recommendations essentially echoed what this work sought to achieve. The campaign was delivered in a way to meet the target audience needs via social media, developed and piloted with the target audience, and integrated with existing local promotion programs. Twitter, as a vehicle for dissemination and as an opportunity to reach new audiences, is endorsed by The Centers for Disease Control and Prevention [71]; however, fundamental challenges remain. Although respondents to our Web-based surveys were similar to those to the household survey, Twitter users are not representative of the general offline population. There are also ethical and privacy issues surrounding social media and Twitter that have not yet been tackled head on by most public health agencies [72]. For example, difficulties may arise in reporting content, as Tweets can be searched, thus increasing the potential for subjects to be identified [73]. It is not clear to what extent publicly available social media data can be regarded as "public," and as such, "concerns over consent, privacy and anonymity do not disappear simply because subjects participate in Web-based social networks; rather, they become even more important" [74,75]. Issues involving informed consent to social media research have also arisen. For example, Kramer et al's [75] work utilizing Facebook caused expressions of concern from publishers over principles of informed consent. A number of bodies are developing guidelines and protocols for corporate use of social media. Indeed the *Journal of Medical Internet Research* has produced a special issue on "Ethics, Privacy, and Legal Issues" [76], but clear guidance is required whether public health research is to harness its full potential.

## Limitations

This study has generated a number of hypotheses that require testing in a larger, definitive trial. However, a number of limitations have been identified from this study. The issue of contamination across phases remains a key methodological concern in social media research. Future research should seek to employ a phase-based pre-post design and analysis with adequate wash-out period. Another potential threat to the validity of this study, and indeed social media research in general, relates to the unrepresentativeness of the Twitter population. Given the limited social media traffic and interaction with the campaign, our findings should be interpreted with caution. Although we targeted some different types of influencers to aid engagement and reach of the campaign, there were limitations in terms of their number of followers and their overall engagement with the campaign, and therefore these findings too should be interpreted with caution.

## Acknowledgments

The intervention was funded by the Medical Research Council, Public Health Intervention Development scheme, and implemented in collaboration with the Northern Ireland Public Health Agency and the largest regional cancer charity in Northern Ireland, Cancer Focus NI. This work was supported by the Centre of Excellence for Public Health (Northern Ireland), a UKCRC Public Health Research Centre of Excellence. RFH was funded by a Career Development Fellowship from the National Institute of Health Research (NIHR). Approval was sought and granted by the Research Ethics Committee of the School of Medicine, Dentistry and Biomedical Sciences, Queen's University Belfast, Northern Ireland (Ref 14/55).

## Conflicts of Interest

None declared.

## Multimedia Appendix 1

Word searches.

[PDF File (Adobe PDF File), 24KB - publichealth_v3i1e14_app1.pdf ]

## Multimedia Appendix 2

Intervention development.

[PDF File (Adobe PDF File), 194KB - publichealth_v3i1e14_app2.pdf ]

## Multimedia Appendix 3

Checklist for Reporting Results of Internet E-Surveys (CHERRIES).

[PDF File (Adobe PDF File), 56KB - publichealth_v3i1e14_app3.pdf ]

**Multimedia Appendix 4**

Survey advertisement on Twitter.

[[PDF File (Adobe PDF File), 160KB](#) - publichealth_v3i1e14_app4.pdf ]

**References**

1.  Kaplan A, Haenlein M. Users of the world, unite! The challenges and opportunities of social media. Bus Horiz 2010 Jan;53(1):59-68. [doi: [10.1016/j.bushor.2009.09.003](#)]
2.  Syred J, Naidoo C, Woodhall SC, Baraitser P. Would you tell everyone this? Facebook conversations as health promotion interventions. J Med Internet Res 2014 Apr 11;16(4):e108 [[FREE Full text](#)] [doi: [10.2196/jmir.3231](#)] [Medline: [24727742](#)]
3.  Perrin A. Pew Research Center. Social Networking Usage URL: [http://www.pewinternet.org/2015/10/08/social-networking-usage-2005-2015/](#) [accessed 2017-03-02] [[WebCite Cache ID 6ofC9xmEt](#)]
4.  Twitter. URL: [https://about.twitter.com/company](#) [accessed 2017-03-02] [[WebCite Cache ID 6ofCCQPbz](#)]
5.  Facebook. Facebook usage URL: [http://newsroom.fb.com/company-info/](#) [accessed 2017-03-02] [[WebCite Cache ID 6ofCHq9ot](#)]
6.  Duggan M, Ellison N, Lampe C, Lenhart A, Madden M. Pew Research Center. 2016. Social Media Update 2014 URL: [http://www.pewinternet.org/2015/01/09/social-media-update-2014/](#) [accessed 2017-03-02] [[WebCite Cache ID 6ofCKuc9j](#)]
7.  Public Health England Social Marketing Strategy 2014-2017: One year on. -05-06. 2016. July 2015 URL: [https://www.gov.uk/government/uploads/system/uploads/attachment_data/file/445524/Marketing_report_web.pdf](#) [accessed 2017-03-02] [[WebCite Cache ID 6ofCMqkU7](#)]
8.  Schein R, Wilson K, Keelan J. A report for Peel public health. Literature review on effectiveness of the use of social media URL: [https://www.peelregion.ca/health/resources/pdf/socialmedia.pdf](#) [accessed 2017-03-02] [[WebCite Cache ID 6ofCS9ZUv](#)]
9.  Kietzmann J, Hermkens K, McCarthy I, Silvestre B. Social media? Get serious! Understanding the functional building blocks of social media. Bus Horiz 2011 May;54(3):241-251. [doi: [10.1016/j.bushor.2011.01.005](#)]
10. 2012. NHS UK. Using social media to engage, listen and learn. Smart Guides to Engagement URL: [https://www.networks.nhs.uk/nhs-networks/smart-guides/documents/Using%20social%20media%20to%20engage-%20listen%20and%20learn.pdf](#) [accessed 2017-03-02] [[WebCite Cache ID 6ofCfUbsp](#)]
11. Deller R, Tilton S. Selfies as charitable meme: charity and notational identity in the #nomakeupselfie and #thunbsupforstephen campaign. Int J Comm 2015;9:1788-1805.
12. George D, Rovniak L, Kraschnewski J. Dangers and opportunities for social media in medicine. Clin Obstet Gynecol 2013;56(3). [doi: [10.1097/GRF.0b013e318297dc38](#)]
13. Neiger B, Thackery R, Burton S, Giraud-Carrier C, Fagen M. Evaluating social media's capacity to develop engaged audiences in health promotion settings: use of Twitter metrics as a case study. Health Promot Pract 2013;14(2):157-162. [doi: [10.1177/1524839912469378](#)]
14. Chou W, Prestin A, Lyons C, Wen K. Web 2.0 for health promotion: reviewing the current evidence. Am J Public Health 2013 Jan;103(1):e9-e18. [doi: [10.2105/AJPH.2012.301071](#)]
15. Heldman A, Schindelar J, Weaver IJ. Social media engagement and public health communication: implications for public health organizations being truly "social". Public Health Rev 2013;35(1):1.
16. Griffiths F, Lindenmeyer A, Powell J, Lowe P, Thorogood M. Why are health care interventions delivered over the internet? A systematic review of the published literature. J Med Internet Res 2006 Jun 23;8(2):e10 [[FREE Full text](#)] [doi: [10.2196/jmir.8.2.e10](#)] [Medline: [16867965](#)]
17. Thackeray R, Neiger B, Hanson C, McKenzie J. Enhancing promotional strategies within social marketing programs: use of Web 2.0 social media. Health Promot Pract 2008 Oct;9(4):338-343. [doi: [10.1177/1524839908325335](#)]
18. Korda H, Itani Z. Harnessing social media for health promotion and behavior change. Health Promot Pract 2013 Jan;14(1):15-23. [doi: [10.1177/1524839911405850](#)] [Medline: [21558472](#)]
19. Pagoto S, Waring M, May C, Ding E, Kunz W, Hayes R, et al. Adapting behavioral interventions for social media delivery. J Med Internet Res 2016 Jan 29;18(1):e24 [[FREE Full text](#)] [doi: [10.2196/jmir.5086](#)] [Medline: [26825969](#)]
20. Milton N. Nickmilton. 2014. Why knowledge transfer through discussion is 14 times more effective than writing URL: [http://www.nickmilton.com/2014/10/why-knowledge-transfer-through.html](#) [accessed 2017-03-02] [[WebCite Cache ID 6ofDiUH5u](#)]
21. Kofinas J, Varrey A, Sapra K, Kanj R, Chervenak F, Asfaw T. Adjunctive social media for more effective contraceptive counseling: a randomized controlled trial. Obstet Gynecol 2014;123(4):763-770. [doi: [10.1097/AOG.0000000000000172](#)]
22. Schnitzler K, Davies N, Ross F, Harris R. Using Twitter™ to drive research impact: a discussion of strategies, opportunities and challenges. Int J Nurs Stud 2016 Jul;59:15-26. [doi: [10.1016/j.ijnurstu.2016.02.004](#)]
23. Laranjo L, Arguel A, Neves A, Gallagher A, Kaplan R, Mortimer N, et al. The influence of social networking sites on health behavior change: a systematic review and meta-analysis. J Am Med Inform Assoc 2014 Jul 08;22(1):243-256. [doi: [10.1136/amiajnl-2014-002841](#)]

24. Signorini A, Segre AM, Polgreen PM. The use of Twitter to track levels of disease activity and public concern in the U.S. during the influenza A H1N1 pandemic. PLoS One 2011 May 04;6(5):e19467 [FREE Full text] [doi: 10.1371/journal.pone.0019467] [Medline: 21573238]
25. Lefebvre RC, Bornkessel AS. Digital social networks and health. Circulation 2013 Apr 30;127(17):1829-1836 [FREE Full text] [doi: 10.1161/CIRCULATIONAHA.112.000897] [Medline: 23630086]
26. Department for Health, Social Services and Public Safety (DHSSPS). Skin cancer prevention - strategy and action plan (2011-2021). Department of Health, UK 2011 [FREE Full text]
27. Donnelly DW, Gavin AT. Northern Ireland Cancer Registry. 2015. Cancer incidence trends 1993-2013 with projections to 2035 URL: https://www.qub.ac.uk/research-centres/nicr/FileStore/PDF/NIrelandReports/Filetoupload,531911,en.pdf [accessed 2017-03-08] [WebCite Cache ID 6oo1HXctO]
28. Montague M, Borland R, Sinclair C. Slip! Slop! Slap! and SunSmart, 1980-2000: Skin cancer control and 20 years of population-based campaigning. Health Educ Behav 2001 Jun;28(3):290-305. [doi: 10.1177/109019810102800304] [Medline: 11380050]
29. Miles A, Waller J, Hiom S, Swanston D. SunSmart? Skin cancer knowledge and preventive behaviour in a British population representative sample. Health Educ Res 2005 Oct;20(5):579-585 [FREE Full text] [doi: 10.1093/her/cyh010] [Medline: 15644381]
30. Eysenbach G. CONSORT-EHEALTH: improving and standardizing evaluation reports of Web-based and mobile health interventions. J Med Internet Res 2011 Dec 31;13(4):e126 [FREE Full text] [doi: 10.2196/jmir.1923] [Medline: 22209829]
31. Lister C, Royne M, Payne HE, Cannon B, Hanson C, Barnes M. The Laugh Model: reframing and rebranding public health through social media. Am J Public Health 2015 Nov;105(11):2245-2251. [doi: 10.2105/AJPH.2015.302669] [Medline: 26378824]
32. Eysenbach G. Improving the quality of Web surveys: the Checklist for Reporting Results of Internet E-Surveys (CHERRIES). J Med Internet Res 2004 Sep 29;6(3):e34 [FREE Full text] [doi: 10.2196/jmir.6.3.e34] [Medline: 15471760]
33. Carroll C, Booth A, Leaviss J, Rick J. "Best fit" framework synthesis: refining the method. BMC Med Res Methodol 2013;13(1):1. [doi: 10.1186/1471-2288-13-37]
34. Lee EC, Whitehead AL, Jacques RM, Julious SA. The statistical interpretation of pilot trials: should significance thresholds be reconsidered? BMC Med Res Methodol 2014 Mar 20;14(1):41 [FREE Full text] [doi: 10.1186/1471-2288-14-41] [Medline: 24650044]
35. NISRA.: Northern Ireland Statistics and Research Agency; 2012. Census 2011: Key Statistics for Northern Ireland URL: http://www.nisra.gov.uk/Census/key_stats_bulletin_2011.pdf [accessed 2017-03-02] [WebCite Cache ID 6ofGnRk0e]
36. NISRA. Northern Ireland Statistics and Research Agency; 2012. Census 2011: Population and Household Estimates for Local Government Districts in Northern Ireland URL: http://www.nisra.gov.uk/Census/pop_stats_bulletin_2_2011.pdf [accessed 2017-03-02] [WebCite Cache ID 6ofGsX8mm]
37. Hatch EE, Hahn KA, Wise LA, Mikkelsen EM, Kumar R, Fox MP, et al. Evaluation of selection bias in an internet-based study of pregnancy planners. Epidemiology 2016 Jan;27(1):98-104 [FREE Full text] [doi: 10.1097/EDE.0000000000000400] [Medline: 26484423]
38. Hollier L, Pettigrew S, Slevin T, Strickland M, Minto C. Comparing online and telephone survey results in the context of a skin cancer prevention campaign evaluation. J Public Health 2016;10:1-9. [doi: 10.1093/pubmed/fdw018]
39. Gallagher K, Updegraff J. Health message framing effects on attitudes, intentions, and behavior: a meta-analytic review. Ann Behav Med 2011 Oct 13;43(1):101-116. [doi: 10.1007/s12160-011-9308-7]
40. O'Keefe D, Wu D. Gain-framed messages do not motivate sun protection: a meta-analytic review of randomized trials comparing gain-framed and loss-framed appeals for promoting skin cancer prevention. Int J Environ Res Public Health 2012 Jun 05;9(12):2121-2133. [doi: 10.3390/ijerph9062121]
41. Hammond D. Health warning messages on tobacco products: a review. Tob Control 2011;20:327-337 [FREE Full text] [doi: 10.1136/tc.2010.037630]
42. Mays D, Niaura R, Evans W, Hammond D, Luta G, Tercyak K. Cigarette packaging and health warnings: the impact of plain packaging and message framing on young smokers. Tob Control 2014 [FREE Full text] [doi: 10.1136/tobaccocontrol-2013-051234]
43. Dunlop S, Wakefield M, Kashima Y. Can you feel it? Negative emotion, risk, and narrative in health communication. Media Psychol 2008 Mar 21;11(1):52-75. [doi: 10.1080/15213260701853112]
44. Paek H, Oh S, Hove T. How fear-arousing news messages affect risk perceptions and intention to talk about risk. Health Commun 2016;31(9):1051-1062. [doi: 10.1080/10410236.2015.1037419]
45. Witte K, Allen M. A meta-analysis of fear appeals: implications for effective public health campaigns. Health Educ Behav 2000;27(5):591-615. [doi: 10.1177/109019810002700506]
46. Leshner G, Bolls P, Thomas E. Scare' em or disgust 'em: the effects of graphic health promotion messages. Health Commun 2009 Jul;24(5):447-458. [doi: 10.1080/10410230903023493] [Medline: 19657827]
47. Cho H, Salmon CT. Unintended effects of health communication campaigns. J Commun 2007 Jun;57(2):293-317. [doi: 10.1111/j.1460-2466.2007.00344.x]

XSL•FO
RenderX

48. Bail CA. Emotional feedback and the viral spread of social media messages about Autism spectrum disorders. Am J Public Health 2016 Jul;106(7):1173-1180. [doi: 10.2105/AJPH.2016.303181] [Medline: 27196641]
49. Ramdhani I. NTUST. Motives of twitter users to retweet URL: http://pc01.lib.ntust.edu.tw/ETD-db/ETD-search/view_etd?URN=etd-0716112-105845 [accessed 2017-03-03] [WebCite Cache ID 6ogjdgf0k]
50. Kandadai V, Yang H, Jiang L, Yang C, Fleisher L, Winston F. Measuring health information dissemination and identifying target interest communities on Twitter: methods development and case study of the @SafetyMD network. JMIR Res Protoc 2016 May 05;5(2):e50. [doi: 10.2196/resprot.4203]
51. Brett B. The New York Times. The psychology of sharing URL: http://nytmarketing.whsites.net/mediakit/pos/ [accessed 2017-03-03] [WebCite Cache ID 6ogjj4uCI]
52. The New York Times. Social Bakers, 2015 URL: http://nytmarketing.whsites.net/mediakit/pos/ [accessed 2017-03-03] [WebCite Cache ID 6ogjjrkDq]
53. Bakshy E, Hofman J, Mason W, Watts D. Everyone's an influencer: quantifying influence on twitter. 2011 Presented at: InProceedings of the fourth ACM international conference on Web searchdata mining; February 9, 2011; Hong Kong, China p. 65-74. [doi: 10.1145/1935826.1935845]
54. Romero D, Galuba W, Asur S, Huberman B. Influence and passivity in social media. In: Machine learning and knowledge discovery in databases. Berlin, Heidelberg: Springer-Verlag; 2011:18-33.
55. Remy C, Pervin N, Toriumi F, Takeda H. Information Diffusion on Twitter: Everyone has its chance, but all chances are not equal. 2013 Dec 05 Presented at: Signal-Image Technology & Internet-Based Systems (SITIS), International Conference on Dec 2 (pp. ). IEEE; 2013; Washington, DC p. 483-490. [doi: 10.1109/SITIS.2013.84]
56. Yang J, Counts S. Predicting the Speed, Scale, and Range of Information Diffusion in Twitter. 2010 May 10 Presented at: ICWSM; May 10, 2010; Washington, DC p. 10-18.
57. Suh B, Hong L, Pirolli P, Chi E. Want to be retweeted? large scale analytics on factors impacting retweet in twitter network. In: SOCIALCOM '10 Proceedings of the 2010 IEEE Second International Conference on Social Computing. 2010 Presented at: IEEE Second International Conference on Social Computing; August 20, 2010; Washington, DC p. 177-184. [doi: 10.1109/SocialCom.2010.33]
58. Naveed N, Gottron T, Kunegis J, Alhadi A. Bad news travel fast: A content-based analysis of interestingness on twitter. 2011 Jun 15 Presented at: The 3rd International Web Science Conference; June 15, 2011; Koblenz, Germany.
59. Wang F, Wang H, Xu K. Diffusive logistic model towards predicting information diffusion in online social networks. 2012 Presented at: 2012 32nd International Conference on Distributed Computing Systems Workshops; June 18, 2012; Macau, China p. 133-139. [doi: 10.1109/ICDCSW.2012.16]
60. Hong L, Dan O, Davison B. Predicting popular messages in twitter. 2011 Presented at: 20th international conference companion on World wide web; March 28, 2011; Hyderabad, India p. 57-58.
61. Yang J, Leskovec J. Modeling information diffusion in implicit networks. 2010 Dec Presented at: IEEE 10th International Conference on Data Mining (ICDM); 2010; University of Technology Sydney, Australia p. 599-608. [doi: 10.1109/ICDM.2010.22]
62. Moz Blog.: Moz Blog; 2012. When Is My Tweet's Prime of Life? (A brief statistical interlude) URL: https://moz.com/blog/when-is-my-tweets-prime-of-life [accessed 2016-05-06] [WebCite Cache ID 6oglpXN2C]
63. Cobb N, Jacobs M, Wileyto P, Valente T, Graham A. Diffusion of an evidence-based smoking cessation intervention through Facebook: a randomized controlled trial. Am J Public Health 2016;106(6):1130-1135. [doi: 10.2105/AJPH.2016.303106]
64. Tufekci Z. Big questions for social media big data: Representativeness, validity and other methodological pitfalls. In: ICWSM '14: Proceedings of the 8th International AAAI Conference on Weblogs and Social Media, 2014. 2014 Mar 28 Presented at: 8th International AAAI Conference on Weblogs and Social Media; 2014; Ann Arbor, Michigan.
65. Wu YP, Aspinwall LG, Conn BM, Stump T, Grahmann B, Leachman SA. A systematic review of interventions to improve adherence to melanoma preventive behaviors for individuals at elevated risk. Prev Med 2016 Jul;88:153-167. [doi: 10.1016/j.ypmed.2016.04.010]
66. Kozinets R. The field behind the screen: using netnography for marketing research in online communities. J Mark Res 2002;39(1):61-72.
67. Hannak A, Anderson E, Barrett L, Lehmann S, Mislove A, Riedewald M. Tweetin' in the Rain: Exploring Societal-Scale Effects of Weather on Mood. In: Sixth International AAAI Conference on Weblogs and Social Media. 2012 Presented at: Sixth International AAAI Conference on Weblogs and Social Media; June 4–7, 2012; Dublin, Ireland.
68. Wood I. The mixed emotions of Twitter. 2016 Presented at: Twitter for Research Conference; April 20, 2016; Galway, Ireland.
69. Freeman B, Potente S, Rock V, McIver J. Social media campaigns that make a difference: what can public health learn from the corporate sector and other social change marketers? Public Health Res Pract 2015;25(2).
70. NICE. 2016. NICE Guidelines NG34 Sunlight exposure: risks and benefits, February 2016 URL: https://www.nice.org.uk/guidance/ng34 [accessed 2017-03-03] [WebCite Cache ID 6ogmU9gcW]
71. Centers for Disease Control and Prevention. Atlanta Social media at CDC: Twitter URL: https://www.cdc.gov/socialmedia/tools/twitter.html [accessed 2017-03-03] [WebCite Cache ID 6ogmgWVhT]

72. McKee R. Ethical issues in using social media for health and health care research. Health Policy 2013 May;110(2-3):298-301. [doi: 10.1016/j.healthpol.2013.02.006] [Medline: 23477806]
73. Zimmer M. "But the data is already public": on the ethics of research in Facebook. Ethics Inf Technol 2010 Jun 4;12(4):313-325. [doi: 10.1007/s10676-010-9227-5]
74. Spiro E. Research opportunities at the intersection of social media and survey data. Curr Opin Psychol 2016 Jun;9:67-71. [doi: 10.1016/j.copsyc.2015.10.023]
75. Kramer AD, Guillory JE, Hancock JT. Experimental evidence of massive-scale emotional contagion through social networks. In: Proceedings of the National Academy of Sciences. 2014 Presented at: Proceedings of the National Academy of Sciences of the United States of America, PNAS; June 17, 2014; USA p. 8788-8790. [doi: 10.1073/pnas.1320040111]
76. Journal of Medical Internet Research. Journal of Medical Internet Research. E-collection: Ethics, Privacy, and Legal Issues. Journal of Medical Internet Research URL: http://www.jmir.org/themes/37 [accessed 2017-03-03] [WebCite Cache ID 6ogrYrLps]

## Abbreviations

**UV:** Ultraviolet
**CHERRIES:** Checklist for Reporting Results of Internet E-Surveys